\documentclass[preprint]{aastex}
\usepackage{graphicx}
\usepackage{dcolumn}
\usepackage{bm}
\usepackage{hyperref}
\usepackage{subfigure}
\begin{document}

\title{O-V-S-Z and friends: Non-Gaussianity from inhomogeneous reionization}
\author{Rishi Khatri}
\email{rkhatri2@illinois.edu}
\affil{Department of Astronomy, University of Illinois at Urbana-Champaign, 1002 W.~Green Street, Urbana, IL 61801}
\and
\author{Benjamin D. Wandelt\altaffilmark{1,2}}
\email{bwandelt@illinois.edu}
\affil{Department of Physics, University of Illinois at Urbana-Champaign,
  1110 W.~Green Street, Urbana, IL 61801, USA}
\altaffiltext{1}{Department of Astronomy, University of Illinois at
  Urbana-Champaign, 1002 W.~Green Street, Urbana, IL 61801, USA}
\altaffiltext{2}{California Institute of Technology, Mail Code 130-33,
  Pasadena, CA 91125, USA}
\date{\today}
\begin{abstract}
We calculate the cosmic microwave background (CMB) bispectrum due to
inhomogeneous reionization. We calculate all the terms that can contribute
to the bispectrum that are products of first order terms on all
scales in conformal Newtonian gauge. We also correctly account for the de-correlation between the matter
density and initial conditions using perturbation theory up to third order.
We find that the bispectrum is of local type as expected.
 For a reasonable model of reionization, in which the universe
 is completely ionized by redshift $z_{ri}\sim 8$ with optical depth to the
 last scattering surface $\tau_0=0.087$ the signal-to-noise ratio for detection of the CMB temperature bispectrum is $\rm{S/N}\sim 0.1$ and
 confusion in the estimation of primordial non-Gaussianity is  $f_{NL}\sim
 -0.1$. For an extreme model with $z_{ri}\sim 12.5$ and $\tau_0=0.14$ we get
 $\rm{S/N}\sim 0.5$ and $f_{NL}\sim -0.2$.
 \end{abstract}
\maketitle
\begin{section}{Introduction}
Secondary anisotropies \citep{review} in the cosmic microwave background CMB can be used to probe the universe after
recombination. It is also important to take them into account when using
CMB to learn about the initial conditions of the universe. One important
class of secondary anisotropies arises due to the scattering of CMB photons by
free electrons during and after reionization. In this class,  cosmologists have
so far concentrated on only one of the terms in the second order Boltzmann
equation, the product of electron velocity and electron number
density ($v_e n_e$). It is known as Sunyaev-Zel'dovich effect when the source of
electrons is hot gas in galaxy clusters \citep{sz}. If instead of thermal
motion velocity due to the bulk
motion of electrons is considered it is known as Ostriker-Vishniac (OV) effect
\citep{ov,vish2} or kinetic Sunyaev-Zel'dovich  (kSZ) effect. There are however
additional terms in the full second order equations \citep{bar1,bar,pitrou,sen1} which also
arise due to scattering of CMB photons by electrons and which might be
important. Most of the work on OV/kSZ effect has focused on the CMB power
spectrum . The CMB bispectrum and trispectrum were calculated in
\citet{castro,castro2}, however they calculated the next to leading order term
which is a six point correlation function of first order terms for the bispectrum. The
leading order term in bispectrum is a four point correlation function of first order
terms. They also ignored the de-correlation between the linear and the non-linear
quantities in their calculation.

The CMB bispectrum due to inhomogeneous recombination was calculated in
\citet{kw1} hereafter referred to as KW09 (see \citet{sen2} for a
different approach, also \citet{kw2}). The same equations need to be solved
for inhomogeneous reionization and we will follow the treatment in
KW09. We will model the inhomogeneous reionization using the linear
perturbation theory
of  \citet{lri} (hereafter ZHH07). For the recombination case the Doppler terms which give rise to
OV/kSZ effects were found to be sub-dominant compared to the net contribution
from the quadrupole and higher order moments of the CMB. We will see that
this is also the case for reionization. For all calculations the gauge
dependent quantities are in conformal Newtonian gauge. The cosmological
parameters used are baryon density $\Omega_b=0.048$, cold dark matter
density $\Omega_c=0.252$, cosmological constant $\Omega_{\Lambda}=0.7$,
number of massless neutrinos $N_{\nu}=3.04$, Hubble constant $H_0=69$,
present CMB temperature $T_{CMB}=2.725$, primordial Helium fraction
$y_{He}=0.24$, spectral index of primordial fluctuations $n_s=0.95$ and $\sigma_8=0.826$.
\end{section}

\begin{figure}
\includegraphics{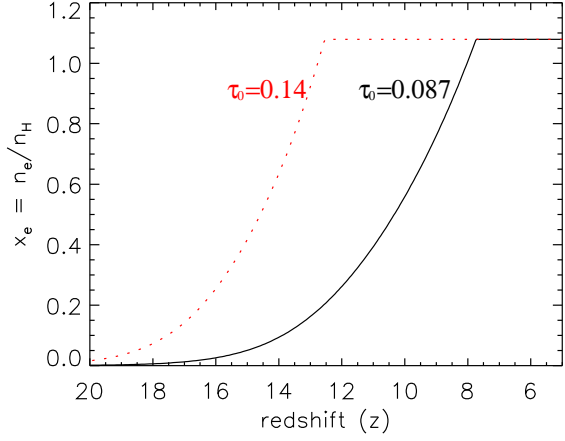}
\caption{\label{xe}Reionization history for two models with optical
  depth to the last scattering surface $\tau_0= 0.087,0.14$.}
\end{figure}

\begin{figure}
\includegraphics{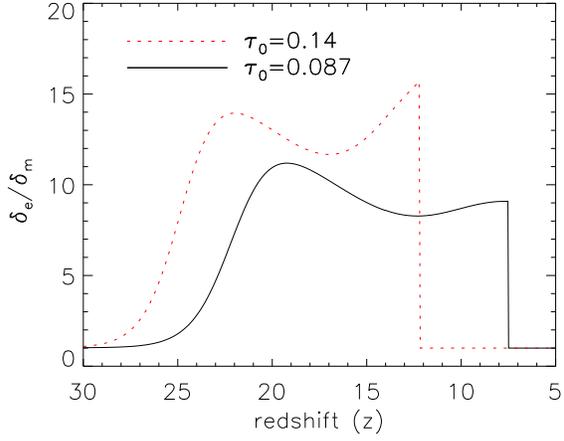}
\caption{\label{dxe}$b_e\equiv\delta_e/\delta_m$ for $k=0.01 {\rm Mpc}^{-1}$ and  optical
  depth to the last scattering surface $\tau_0= 0.087, 0.14$.}
\end{figure}

\begin{section}{Inhomogeneous reionization}
We will use the linear perturbation theory of ZHH07 to model reionization. The results from this model are
similar to the bubble model of reionization \citep{bubble}. Due to the fact
that reionization is sourced by non-linear physics, the validity of any
model will have to be tested with computer simulations (see \citet{rirev}
for a recent review). For our purpose the analytical treatment of
ZHH07, which captures the
essential features of reionization on linear scales, is sufficient. We refer the reader to
ZHH07 for details of the model as well as for discussion on the
validity of this approach. An important input for this theory is a model
for the distribution of ionizing sources. We will use the same model used
in ZHH07  which is based on the excursion set treatment of halo formation
\citep{ps,bond,lc} with the minimum mass of a halo given by virial temperature of $10^4$K
 corresponding to where  hydrogen line cooling becomes efficient. The
 spectrum of ionizing radiation is taken to be a power law
\begin{eqnarray}
\gamma (\mu) d\mu &=& \frac{\zeta}{C_{\beta}}e^{(\beta +1)\mu}d\mu\label{gmu}
\end{eqnarray}
where $\mu=\ln \nu - \ln \nu_0$, $\nu$ is the photon frequency,
$\nu_0=13.6{\rm eV}/2\pi\hbar$ is the ionization threshold for hydrogen,
$\hbar$ is the Planck's constant, $\gamma(\mu)$ is the number of ionizing
photons emitted at frequency $\nu$ per unit parameter $\mu$  per collapsed hydrogen atom, $\beta$ is
the spectral index of ionizing radiation spectrum, $\zeta$ is the total
number of ionizing photons emitted per hydrogen atom and $C_{\beta}=\int_0^{10}e^{\left(\beta +1\right)\mu}d\mu$ is the
normalization constant with the spectrum cutoff at $\mu=10$. We take into
account Helium reionization by assuming that the first ionization of Helium
is identical to that of Hydrogen. Although not strictly correct, it should
introduce only a small error, unimportant for us, since Helium will
contribute only about $8\%$ of the total electrons. Second ionization of Helium
is expected to occur at much lower redshifts \citep{furhe} and will give a negligible
contribution to the CMB bispectrum. We will consider two different models of
reionization arrived at by choosing different values of parameter $\zeta$
in Equation \ref{gmu} with spectral index $\beta=-3$. For the first model we
choose $\zeta=70$ to give the optical depth to the last scattering surface
 $\tau_0=0.087$. For the second model we choose $\zeta=1000$ resulting in
$\tau_0=0.14$ which can be considered
a reasonable upper limit based on WMAP 5 year results \citep{wmap5}. Figure
\ref{xe} shows the reionization history for these two models. We use the
RECFAST code \citep{recfast} to calculate the residual mean electron
number density after recombination switching to reionization code once the
electron density due to reionization exceeds the residual value from recombination. The ratio of electron number density perturbation to matter density
perturbation for comoving wavenumber of $k=0.01 {\rm Mpc}^{-1}$ is plotted
in Figure \ref{dxe}. We use the approximate solutions to the perturbation
equations given in  ZHH07 and force the electron bias $b_e\equiv\delta_e/\delta_m=1$ once
the universe is fully reionized. This is a very good approximation to the
exact equations of ZHH07 where the bias $b_e$ goes smoothly to unity. The matter density and hence the electron number
density will be non-linear on small scales and thus will de-correlate with
the linear quantities, for example CMB, on these scales. We will use $\delta_e=b_e \delta_m$
on all scales, where $b_e$ is
calculated using linear theory but $\delta_m$ can be non-linear. We will take the
de-correlation into account using the third order perturbation theory. Note
that the non-linearity will be significant only at low redshifts for scales
of interest when the universe is fully reionized and $\delta_e=\delta_m$
exactly. Also for the leading term in the  bispectrum we need to correlate CMB
with the electron number density perturbation. However  this correlation
will be small  on scales much smaller than
the horizon size because CMB traces the perturbations at a much higher
redshift than that of reionization. Thus  the contributions to the bispectrum from
perturbations in the electron number density will be significant only for
scales which are linear and where we should expect the linear perturbation
theory of reionization to work well. 

At low redshifts a significant
fraction of baryons are expected  to be in a diffuse phase called
warm-hot intergalactic medium (WHIM) \citep{cen1,cen2}. The perturbations
in these baryons are suppressed on small (non-linear) scales compared to
the dark matter and for these
baryons the bias $b_e$ should be less than $1$. However 
the contribution to the bispectrum from $3 <z < 6$ is $\sim
10\%$ and contribution from $z<3$ is $\sim \rm{few}\%$. This is because of the
absence of bias $b_e \sim 10$ due
to the inhomogeneities in the reionization process once the universe is
completely reionized and the decreasing
optical depth due to the expansion of the universe. At $3 < z < 6$ only
a small percentage of baryons are in WHIM ($< 10\%$). Thus their contribution to the bispectrum
is less than $1\%$ and the error in assuming $b_e = 1$ negligible. We neglect the $\sim \rm{few}\%$ contribution
from $z < 3$ in our numerical calculations.
\end{section}

\begin{section}{Cross correlation between matter density and initial
    conditions}
We will need to calculate the correlation ($P_X$) between the
electron number density and linear perturbation variables or equivalently the
cross-correlation
between the non-linear matter density ($\delta_m$) and linearly evolved
matter density ($\delta_L$).
We will be interested in the CMB anisotropies on scales of angular wavenumber
$\ell \le 2500$, corresponding to the smallest scales predicted to be
probed by the Planck mission \footnote{\url{http://www.rssd.esa.int/Planck}}. On these scales it is
sufficiently accurate to calculate the next term in the perturbation expansion
which, for $P_X$, means going to the third order in perturbation theory for
matter density
perturbation $\delta_m$. For  Einstein-De Sitter universe
($\Omega_m=1,\Omega_{\Lambda}=0$) the solution for $\delta_m$ can be
written as the following
perturbation series
\citep{vish,goroff,makino,jain} (see \citet{prev} for a review).
\begin{eqnarray}
\delta_m(\mathbf{k},\eta)&=&\sum_{n=1}^{\infty}a^n(\eta)\delta_n(\mathbf{k}),
\end{eqnarray}
where $\delta_1$ is the linear matter density perturbation at $z=0$ and $\delta_n$
is of order  $\delta^n_1$, $\mathbf{k}$ is the Fourier wavenumber and
$\eta$ is conformal time. 
The correlation between the linear and the non-linear matter density is then given by
\begin{eqnarray}
\langle a(\eta)\delta_1(\mathbf{k}) \delta_m(\mathbf{k'},\eta) \rangle &=&
a^2(\eta)\langle \delta_1(\mathbf{k}) \delta_1(\mathbf{k'})
\rangle + a^4(\eta)\langle \delta_1(\mathbf{k}) \delta_3(\mathbf{k'})
\rangle + {\rm higher \hspace{4 pt} order \hspace{4 pt} terms}.\nonumber\\
&=& \left(2\pi\right)^3\delta_D(\mathbf{k+k'})\left[a^2(\eta)P_{11}(k) + a^4(\eta)P_{13}(k)\right],
\end{eqnarray}
where $P_{11}(k)$ is the linear matter power spectrum at $z=0$ and $P_{13}$ is
the correction given by \citep{suto}
\begin{eqnarray}
P_{13}(k)&=&\frac{2\pi
  k^2}{504}P_{11}(k)\int_0^{\infty}\frac{dq}{(2\pi)^3}P_{11}(q)\nonumber\\
&&\times\left[12\frac{k^2}{q^2}-158+100\frac{q^2}{k^2}-42\frac{q^4}{k^4}+3\frac{k^3}{q^3}\left(\frac{q^2}{k^2}-1\right)^3\left(7\frac{q^2}{k^2}+2\right)\ln\left(\frac{k+q}{|k-q|}\right)\right]
\end{eqnarray}
For a general cosmology replacing the scale
factor $a(\eta)$ with the linear growth factor $D(\eta)$ in the Einstein
De-Sitter solution gives an
excellent approximation to the true result \citep{sco,prev}. Thus we have
\begin{eqnarray}
\langle \delta_L(\mathbf{k},\eta) \delta_m(\mathbf{k'},\eta) \rangle
&=&\left(2\pi\right)^3\delta_D(\mathbf{k+k'})P_X(k,\eta),\nonumber\\
P_X(k,\eta)&=&D^2(\eta)P_{11}(k)+D^4(\eta)P_{13}(k),
\end{eqnarray}
where $P_{13}(k)$ is negative signifying de-correlation between the linear and
the non-linear density fields as expected.

We plot the ratio of cross power spectrum $P_X$ to linear power spectrum
$P_{lin}(k,\eta)\equiv D^2(\eta)P_{11}(k)$ in Figure \ref{cross}. For
$\ell\le 2500$, the CMB bispectrum will get contributions from Fourier modes
$k\lesssim 0.4{\rm Mpc}^{-1}$.  It is evident from Figure \ref{cross} that for $k\gtrsim 0.1{\rm Mpc}^{-1}$
the matter density perturbations become mildly non-linear
(i.e. $0.75\lesssim P_X/P_{lin}\lesssim 1$). On these scales
comparison with $N$-body simulations shows that going up to third order in
perturbation theory is a very good approximation while on
smaller scales third order perturbation theory underestimates the
cross-correlation between the linear and non-linear matter density
fields \citep{jeong,carl}. 
 Taking this de-correlation into account results in replacing the linear power spectrum
$P_{lin}$ in the bispectrum expression involving $\delta_e$ by the cross power spectrum $P_X$.
 Equivalently we can define an effective transfer
  function that we can use in the bispectrum expressions derived in KW09.
\begin{eqnarray}
 \delta_e^{eff}&\equiv&\delta_e\frac{P_X}{P_{lin}}
\end{eqnarray}
\begin{figure}
\includegraphics{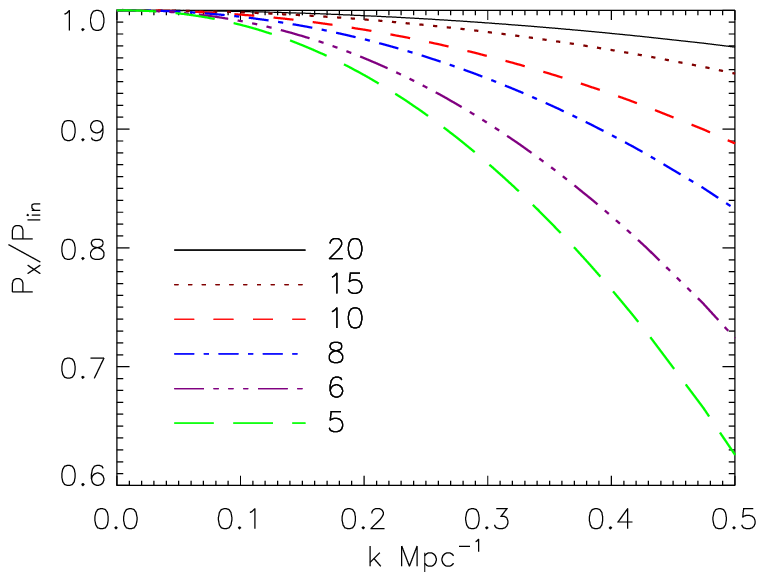}
\caption{\label{cross}Ratio of cross power spectrum between linear and
  non-linear matter density fields to linear matter power spectrum ($P_X/P_{lin}$) for different redshifts.}
\end{figure}

\end{section}

\begin{section}{Bispectrum}
The CMB bispectrum is given by
\begin{eqnarray}
B_{m_1m_2m_3}^{\ell_1\ell_2\ell_3}&=&\langle
a_{\ell_1m_1}^{(1)}(\bm{x},\eta_0)a_{\ell_2m_2}^{(1)}(\bm{x},\eta_0)a_{\ell_3m_3}^{(2)}(\bm{x},\eta_0)\rangle
+ \hspace{4 pt}2\hspace{4 pt}\rm{permutations},
\end{eqnarray}
where $a_{\ell m}(\bm{x},\eta_0)$ is the  spherical harmonic transform
of CMB temperature field and superscript indicates the perturbation order.
Taking into account all terms that multiply $\delta_e$ in the second order
Boltzmann equation for photons results in the following expression for the
angular averaged bispectrum (see KW09 for details of the derivation).
\begin{eqnarray}
B^{\ell_1\ell_2\ell_3}&=&\sum_{m_1m_2m_3}\left(
\begin{array}{lcr}
\ell_1 & \ell_2 & \ell_3 \\
  m_1 & m_2 & m_3
\end{array}
\right)
B_{m_1m_2m_3}^{\ell_1\ell_2\ell_3}\nonumber\\
&=&\sqrt{\frac{(2\ell_1 +1)(2\ell_2 +1)
(2\ell_3 +1)}{4\pi}}\left(
\begin{array}{lcr}
\ell_1 & \ell_2 & \ell_3 \\
  0 & 0 & 0
\end{array}
\right)
\int_0^{\eta_0}d\eta
g(\eta)\biggl[ B_{\delta\Theta}^{\ell_1}(\eta)B_{\Theta\Theta}^{\ell_2}(\eta)\nonumber\\
&&+
B_{\delta\Theta}^{\ell_2}(\eta)B_{\Theta\Theta}^{\ell_1}(\eta)+
B_{\delta\Theta}^{\ell_2}(\eta)B_{\Theta\Theta}^{\ell_3}(\eta)+
B_{\delta\Theta}^{\ell_3}(\eta)B_{\Theta\Theta}^{\ell_2}(\eta)+
B_{\delta\Theta}^{\ell_1}(\eta)B_{\Theta\Theta}^{\ell_3}(\eta)+
B_{\delta\Theta}^{\ell_3}(\eta)B_{\Theta\Theta}^{\ell_1}(\eta)
\biggr]\nonumber\\
\label{blll}\\
B_{\delta\Theta}^{\ell_1}(\eta)&=&\frac{2}{\pi}\int
k_1^2dk_1P(k_1)\Theta_{\ell_1}^{(1)}(k_1,\eta_0)\delta_e^{eff}(k_1,\eta)j_{\ell_1}[k_1(\eta_0-\eta)]\label{bdt}\\
B_{\Theta\Theta}^{\ell_2}(\eta)
&=&\frac{2}{\pi}\int
k_2^2dk_2P(k_2)\Theta_{\ell_2}^{(1)}(k_2,\eta_0)\biggl[\nonumber\\
&&-\sum_{\ell ''\geq 2, \ell_2'} i^{\ell ''+\ell_2 + \ell_2 '}(-1)^{\ell_2}(2\ell
  ''+1)(2\ell_2 '+1)\left(
\begin{array}{lcr}
\ell_2 ' & \ell_2 & \ell '' \\
  0 & 0 & 0
\end{array}
\right)^2 
\Theta_{\ell ''}^{(1)}(k_2,\eta)j_{\ell_2'}[k_2(\eta_0-\eta)]\nonumber\\
& + &\left[\theta_b(k_2,\eta)-\theta_{\gamma}(k_2,\eta)\right]\frac{j'_{\ell_2}
  [k_2(\eta_0-\eta)]}{k_2}\nonumber\\
\label{btt}& +&\frac{1}{4}\Pi^{(1)}(k_2,\eta)\left\{3j''_{\ell_2}[k_2(\eta_0-\eta)]+j_{\ell_2}\left[k_2(\eta_0-\eta)\right]\right\}
\biggr]
\end{eqnarray}
where $g(\eta)$ is the visibility function, $j_{\ell}$ is spherical Bessel
function, $\Theta_{\ell}^{(1)}$ are the first order CMB transfer functions,
with $\Theta^{(n)} \equiv \Delta T^{(n)}/T$ the CMB temperature
perturbation, the matrices are Wigner 3-$jm$ symbols, $\theta_b=ikV_b$, $V_b$
is the baryon velocity, $\theta_{\gamma}=3k\Theta_1^{(1)}$, $\Pi^{(1)}=
\Theta_2^{(1)}+\Theta_{P0}^{(1)}+\Theta_{P2}^{(1)}$, $\Theta_{P\ell}^{(1)}$
are spherical harmonic transform coefficients of the polarization field.
$P(k)$ is the power spectrum of the initial gravitational potential.

The $\theta_b-\theta_{\gamma}$ term in Equation \ref{btt} is the OV/kSZ term which has
been the focus of extensive research so far. The last term gives negligible
contribution. The $\sum_{\ell''}\Theta_{\ell''}^{(1)}$ term in Equation
\ref{btt} is the new term and it, we will find, dominates over OV/kSZ term.
During recombination also this term was found to  dominate over other terms
in KW09.  Recently Hernandez-Monteagudo and Sunyaev have calculated the
effect of this term for the scattering of CMB photons in the galaxy clusters \citep{suny}.
\end{section}

\begin{section}{Numerical Results}
We use CMBFAST \citep{cmbfast} to calculate all first order quantities. All
gauge dependent first order quantities are in conformal Newtonian
gauge. Figure \ref{figb} shows $B_{\delta\Theta}$ and contributions from
different terms in $B_{\Theta\Theta}$. It is clear that $\sum_{\ell\ge 2}
\Theta_{\ell}^{(1)}$ gives the dominant contribution. Also the OV term has
a sign opposite to that of $\sum_{\ell\ge 2}
\Theta_{\ell}^{(1)}$ term. We cutoff the sum
at $\ell=1500$ which is sufficient for $\eta\lesssim 7500
{\rm Mpc}$. The  contribution from $\eta> 7500
{\rm Mpc}$ ($z \lesssim 3$) to the bispectrum is small
 ($\sim$few $\%$)  because the
visibility function is small as evident from Figure \ref{vis} and also
because the
perturbation in the ionization fraction is zero since the universe is fully
ionized by this time and we neglect it. Figure
\ref{bispec} shows the absolute value of  bispectrum for our two models of
reionization for $\ell_3=200$. The bispectrum is clearly of local type. It has however a
different shape than the primordial bispectrum of local type
parameterized by the parameter $f_{NL}$. The confusion with the estimators
of the primordial
bispectrum of local type can be quantified by using the following
statistic \citep{fast}:
\begin{eqnarray}
S_{ri}&\equiv&\sum_{\ell_1\leq \ell_2 \leq \ell_3}\frac{B^{\ell_1 \ell_2
    \ell_3}B^{\ell_1 \ell_2
      \ell_3}_{prim}}{C_{\ell_1}C_{\ell_2}C_{\ell_3}}\label{fnl}
  \simeq f_{NL}\sum_{\ell_1\leq \ell_2 \leq \ell_3}\frac{(B^{\ell_1 \ell_2
    \ell_3}_{prim})^2}{C_{\ell_1}C_{\ell_2}C_{\ell_3}}
\end{eqnarray}
Solving Equation \ref{fnl} for $f_{NL}$ gives the confusion that can be
expected if the effects of reionization on the bispectrum were
ignored. This is plotted in Figure \ref{fnlfig}. Planck experiment is
expected to have error bars on $f_{NL}$ of $\sim 5$. The confusion due to the
inhomogeneous reionization is much smaller and thus can be safely ignored
while looking for the primordial non-Gaussianity. We also calculate the
signal-to-noise (S/N) for the detection of the bispectrum due to inhomogeneous
reionization \citep{komatsu01}
\begin{eqnarray}
\frac{S}{N} &  \equiv  & \frac{1}{\sqrt{F_{rec}^{-1}}},\nonumber\\
F_{rec} & = & \sum_{\ell_1\leq \ell_2 \leq \ell_3 \leq \ell_{\max}}\frac{(B^{\ell_1 \ell_2
    \ell_3})^2}{\Delta_{\ell_1 \ell_2
    \ell_3}C_{\ell_1}C_{\ell_2}C_{\ell_3}},\nonumber\\
\Delta_{\ell_1 \ell_2
    \ell_3} & \equiv & 1+\delta_{\ell_1 \ell_2}+\delta_{\ell_2 \ell_3}+\delta_{\ell_3 \ell_1}+2\delta_{\ell_1 \ell_2}\delta_{\ell_2 \ell_3}.
\end{eqnarray}
This is plotted in Figure \ref{sn} for our two reionization models and for
primordial non-Gaussianity with $f_{NL}=1$. For normal reionization with
optical depth $\tau_0=0.087$ the S/N for $\ell_{max}=2500$ is $0.1$. For
the extreme case with $\tau_0=0.14$ we get $\rm{S/N}=0.5$, about $20\%$ more than the primordial
bispectrum with $f_{NL}=1$. 
\begin{figure}
\includegraphics{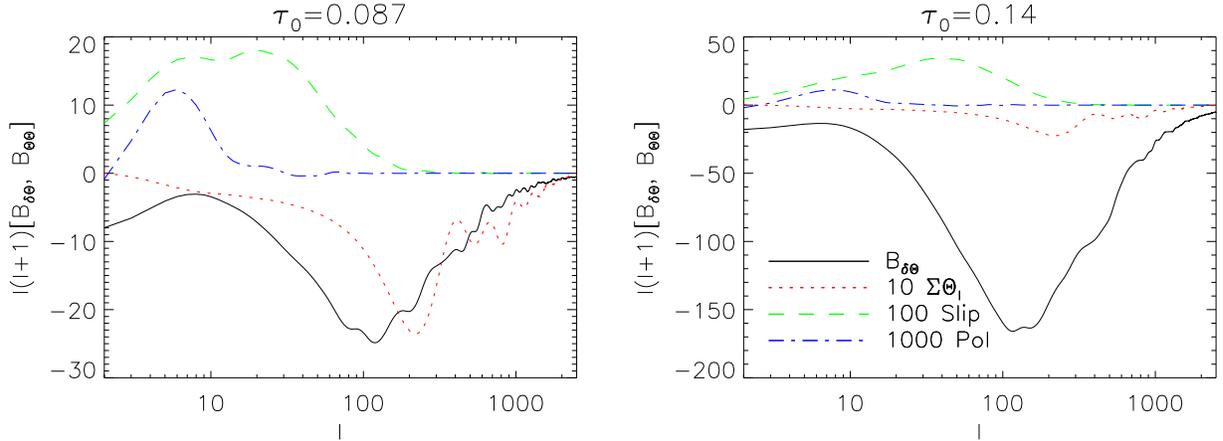}
\caption{\label{figb}$B_{\delta\Theta}$ and contributions to
  $B_{\Theta\Theta}$ from $\sum_{\ell\ge 2}\Theta_{\ell}^{(1)}$, OV term
  $\theta_b-\theta_{\gamma}$ titled ``Slip'' and $\Pi^{(1)}$ titled
  ``Pol'' multiplied by numerical factors as indicated to make them more
  visible. Clearly the dominant contribution to $B_{\Theta\Theta}$ comes
  from the  $\sum_{\ell\ge 2}\Theta_{\ell}^{(1)}$ term. All quantities are
  plotted for $\eta$ near the peak of the visibility function before the
  universe is fully reionized for both models. Note that the sign of OV 
  term is opposite to that of $\sum_{\ell\ge 2}\Theta_{\ell}^{(1)}$ term.}
\end{figure}

\begin{figure}
\includegraphics{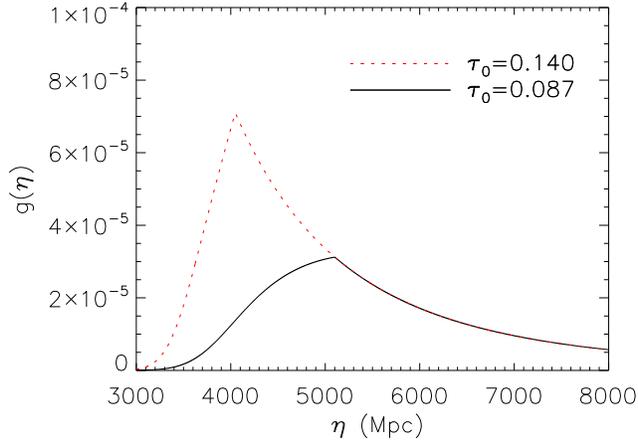}
\caption{\label{vis}Visibility function $g(\eta)$ for our models of reionization}
\end{figure}

\begin{figure}
\includegraphics{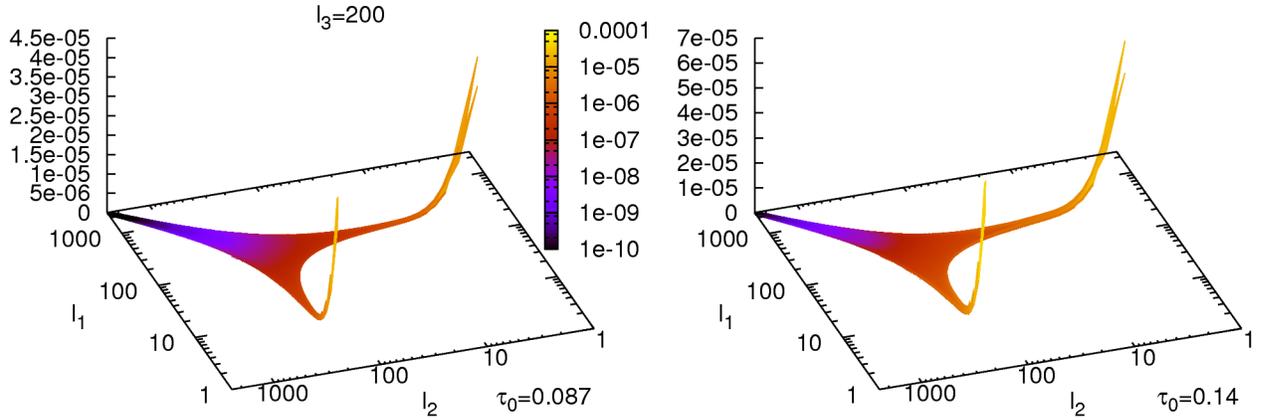}
\caption{\label{bispec}Bispectrum for two models of reionization with
  $\ell_3=200$.The shape of the bispectrum is of local type.}
\end{figure}

\begin{figure}
\includegraphics{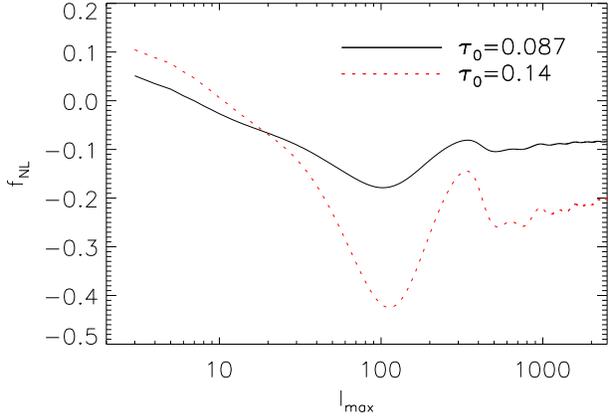}
\caption{\label{fnlfig}Confusion with the primordial non-Gaussianity of
  local type parameterized by $f_{NL}$ (see Equation \ref{fnl}) as a
  function of maximum angular wavenumber $\ell_{max}$.}
\end{figure}

\begin{figure}
\includegraphics{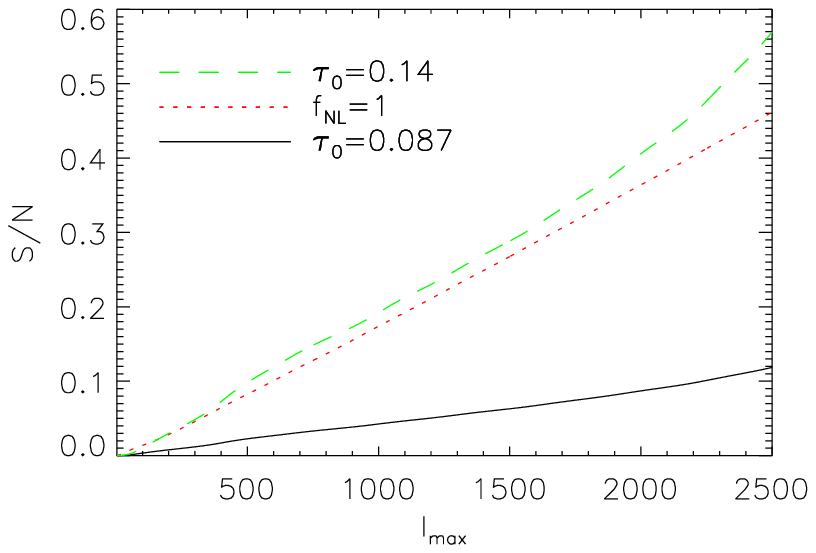}
\caption{\label{sn}S/N for our two
  reionization models assuming a CMB experiment providing a
cosmic variance limited measurement of the anisotropies up to $\ell_{max}$. Also shown for comparison is the S/N for
  the local type primordial non-Gaussianity with $f_{NL}=1$.
}
\end{figure}

\end{section}

\begin{section}{Conclusions}
We have calculated the leading term in the CMB bispectrum due to inhomogeneous reionization.
The bispectrum consists of product of two terms in Equation \ref{blll},
$B^{\ell_i}_{\delta\Theta}B^{\ell_j}_{\Theta\Theta}$, $i,j=1,2,3$.
$B_{\delta\Theta}^{\ell}$ is
due to the correlation of electron number density perturbation with CMB. $B_{\Theta\Theta}^{\ell}$
 is the sum of two
terms, the correlation of CMB with the peculiar velocity of electrons ( the
OV or the kSZ
term) and the correlation of CMB with all higher order moments of CMB. Since CMB traces the
perturbations at a much higher redshift, the correlation of CMB with peculiar velocity in $B_{\Theta\Theta}^{\ell}$
 and
the correlation of CMB with electron number density in 
$B_{\delta\Theta}^{\ell}$ is small on small scales. In particular
the correlation of CMB with CMB in  $B_{\Theta\Theta}^{\ell}$
 dominates over the peculiar velocity or OV/kSZ term.
We have found the bispectrum to be of squeezed triangle type, i.e., it peaks where one $\ell$ mode is
much smaller than the other two with the contribution to the small $\ell$ (large scale) mode coming
from the correlation of $\delta_e$ with CMB and to that of large $\ell$ (small scale) modes coming from the
correlation of CMB with CMB. Note that there will be some correlation of
CMB with $\delta_e$ even on
small scales due to Thomson scattering.

If the correlations of CMB with itself are ignored as has been done prior to this work, the
leading term, which is a four point function of first order terms and which we have calculated,	would be small. In that case the next to leading order term will be a six-point correlation of
only the electron number densities and velocities and may be expected to be
comparatively important
since the electron number density and the velocity would be strongly correlated with each other. This
six point term was calculated in \citet{castro,castro2}. However in the regime where density is
slightly non-linear but velocity is linear, they will get slightly de-correlated. This was ignored
in \citet{castro,castro2} which might have resulted in overestimation of their S/N. They also used an
instantaneous reionization model which does not include the enhancement in the electron number
density perturbation due to inhomogeneous reionization, $b_e$, which is expected to be greater than
one leading to underestimation of their S/N.

The S/N that we get for the leading term including the correlations of CMB
with itself (Figure \ref{sn}) is more than an order of magnitude greater than what was found in  \citet{castro,castro2} for the
next to leading order term. It is still below the detection limit of Planck for
the models considered here.
Thus if the reionization occurs at even higher
redshifts than our extreme model or if the bias $b_e=\delta_e/\delta_m$ is
higher than what the model of reionization used by us predicts, then the imprint of
reionization  in the CMB bispectrum may be seen by Planck or post-Planck
experiments. We would like to point out that there are additional terms in
the second order Boltzmann equation, the second order electron velocity, CMB
monopole and quadrupole, that may also give similar magnitude contributions to the
reionization bispectrum.  The CMB polarization may also get important
contributions from reionization. However the bispectrum is so small, except in the
extreme cases, that it is unlikely that these additional terms would change
our results significantly. More important is the finding that even in
extreme cases the confusion with the primordial non-Gaussianity of local
type is much smaller than one (Figure \ref{fnlfig}). Thus inhomogeneous reionization should not
be a cause of concern when looking for non-Gaussianity in the initial
conditions of the universe in Planck data.
\end{section}
\acknowledgments 
We acknowledge correspondence with Jun Zhang on linear perturbation theory of reionization. We thank Guilhem Lavaux for checking the cross-correlation between the linear and the non-linear matter density perturbations in $N$-body simulation. 
\bibliographystyle{hapj}
\bibliography{ri}
\end{document}